\documentclass[runningheads]{llncs}
\usepackage{graphicx}
\usepackage[utf8]{inputenc}

\usepackage{tikz}
\usepackage{comment}
\usepackage{amsmath,amssymb} 
\usepackage{color}
\usepackage{amsfonts}
\usepackage{booktabs} 
\usepackage{multirow}

\usepackage[implicit=false]{hyperref}
\hypersetup{
    colorlinks=true,
    linkcolor=blue,
    filecolor=magenta,      
    urlcolor=black,
    pdftitle={Overleaf Example},
    pdfpagemode=,
    }

\usepackage[accsupp]{axessibility}  

\begin{document}
\pagestyle{headings}
\mainmatter
\def\ECCVSubNumber{5775}  

\title{Single Frame Atmospheric Turbulence Mitigation: A Benchmark Study and A New Physics-Inspired Transformer Model} 

\titlerunning{ECCV-22 submission ID \ECCVSubNumber} 
\authorrunning{ECCV-22 submission ID \ECCVSubNumber} 
\author{Anonymous ECCV submission}
\institute{Paper ID \ECCVSubNumber}

\titlerunning{Single Frame Turbulence Mitigation}
%
\author{Zhiyuan Mao\inst{1*}\and
Ajay Jaiswal\inst{2*}\and
Zhangyang Wang\inst{2}\and
Stanley H. Chan\inst{1}
}
\authorrunning{Z. Mao et al.}
%
\institute{Purdue University, West Lafayette IN 47907, USA \and
University of Texas at Austin, Austin TX 78712, USA}

\maketitle
\def\thefootnote{*}\footnotetext{Equal contribution.}\def\thefootnote{\arabic{footnote}}
\begin{abstract}
Image restoration algorithms for atmospheric turbulence are known to be much more challenging to design than traditional ones such as blur or noise because the distortion caused by the turbulence is an entanglement of spatially varying blur, geometric distortion, and sensor noise. Existing CNN-based restoration methods built upon convolutional kernels with static weights are insufficient to handle the spatially dynamical atmospheric turbulence effect. To address this problem, in this paper, we propose a physics-inspired transformer model for imaging through atmospheric turbulence. The proposed network utilizes the power of transformer blocks to jointly extract a dynamical turbulence distortion map and restore a turbulence-free image. In addition, recognizing the lack of a comprehensive dataset, we collect and present two new real-world turbulence datasets that allow for evaluation with both classical objective metrics (e.g., PSNR and SSIM) and a new task-driven metric using text recognition accuracy. The code and datasets are available at \href{https://github.com/VITA-Group/TurbNet}{\color{magenta}{github.com/VITA-Group/TurbNet}}.  
\end{abstract}

\section{Introduction}
In long-range imaging systems, atmospheric turbulence is one of the main sources of distortions that causes geometric displacements of the pixels and blurs. If unprocessed, the distorted images can have significant impacts on all downstream computer vision tasks such as detection, tracking, and biometric applications. The atmospheric turbulence effects are substantially harder to model and mitigate compared to the commonly seen image degradations such as deconvolution, as the turbulence is an entanglement of pixel displacement, blur, and noise. As a result, a dedicated image restoration pipeline is an essential element for long-range computer vision problems.

Image processing algorithms for mitigating the atmospheric turbulence effect have been studied for decades \cite{Anantrasirichai2013,mao_tci,Lau2017,Lau2021,Milanfar2013,Yasarla2021ICIP,Hirsch2010,Lou2013,Nair2021ICIP,Li_2021_ICCV}. However, many of them have limitations that prohibit them from being launched to practical systems: 1) Many of the existing algorithms \cite{Anantrasirichai2013,mao_tci,Lau2017,Milanfar2013,Anantrasirichai2013,Hirsch2010} are based on the principle of \emph{lucky imaging} that requires multiple input frames. These methods often have a strong assumption that both the camera and the moving objects are static, which can easily become invalid in many real applications. 2) The conventional algorithms are often computationally expensive, making them unsuitable for processing large-scale datasets to meet the need of the latest computer vision systems. 3) Existing deep learning solutions \cite{Yasarla2021ICIP,Nair2021ICIP,Lau2021} are not utilizing the physics of the turbulence. Many of them are also tailored to recovering faces instead of generic scenes. The generalization is therefore a question. 4) The algorithms may not be properly evaluated due to the absence of a widely accepted real large-scale benchmarking dataset.

To articulate the aforementioned challenges, in this paper we make three contributions:

\begin{enumerate}

\item We present a comprehensive benchmark evaluation of deep-learning based image restoration algorithms through atmospheric turbulence. We tune a sophisticated physics-grounded simulator to generate a large-scale dataset, covering a broad variety of atmospheric turbulence effects. The highly realistic and diverse dataset leads to exposing shortages of current turbulence mitigation algorithms. 

\item Realizing the existing algorithms' limitations, we introduce a novel physics-inspired turbulence restoration model, termed \emph{TurbNet}. Built on a transformer backbone, \emph{TurbNet} features a modularized design that targets modeling the spatial adaptivity and long-range dynamics of turbulence effects, plus a self-supervised consistency loss.

\item We present a variety of evaluation regimes and collect two large-scale real-world turbulence \emph{testing} datasets, one using the heat chamber for classical  objective evaluation (e.g., PSNR and SSIM), and one using real long-range camera for optical text recognition as a semantic ``proxy" task. Both of the new testing sets will be released. 
\end{enumerate}

\section{Related Works}

\textbf{Turbulence mitigation methods. } The atmospheric turbulence mitigation methods have been studied by the optics and vision community for decades. To reconstruct a turbulence degraded image, conventional algorithms \cite{Anantrasirichai2013,mao_tci,Lau2017,Milanfar2013,Hirsch2010,Hardie2017recon,Xie2016,He2016} often adopt the multi-frame image reconstruction strategy. The key idea is called ``lucky imaging", where the geometric distortion is first removed using image registration or optical flow techniques. Sharper regions are then extracted from the aligned frames to form a lucky frame. A final blind deconvolution is usually needed to remove any residue blur. These methods are usually very computationally expensive. The time required to reconstruct a $256\times256$ image may range from a few seconds to tens of minutes. Despite the slow speed that prohibits them from being applied in real-world applications, the performance of conventional methods is often consistent across different image contents. 

Recent deep learning methods adopt more dynamic strategies. Li et al. \cite{Li_2021_ICCV} propose to treat the distortion removal as an unsupervised training step. While it can effectively remove the geometric distortions induced by atmospheric turbulence, its computational cost is comparable to conventional methods, as it needs to repeat the training step for each input image. There are also several works that focus on specific types of images, such as face restoration \cite{Yasarla2021ICIP,Nair2021ICIP,Lau2021}. They are usually based on a simplified assumption on atmospheric turbulence where they assume the blur to be spatially invariant. Such assumption cannot extend to general scene reconstruction, where the observed blur can be highly spatially varying due to a wide field of view. 

There also exists general image processing methods, such as \cite{zamir2021multi,zamir2021restormer}. They have demonstrated impressive performance on restoration tasks, including denoising, deblurring, dehazing, etc. However, whether they can be extended to turbulence mitigation remains unclear as turbulence evolves more complicated distortions. 

\noindent
\textbf{Available datasets. } Despite recent advances in turbulence mitigation algorithms, there is a very limited amount of publicly available datasets for atmospheric turbulence. The most widely used testing data are two images, the \textit{Chimney} and \textit{Building} sequences released in \cite{Hirsch2010}. Besides, authors of \cite{mao_tci,Anantrasirichai2013,Li_2021_ICCV} have released their own testing dataset, each of which often consists of less than 20 images. These data are then seldom used outside the original publications. Additionally, the scale of these datasets is not suitable for evaluating modern learning-based methods. 

Due to the nature of the problem, it is very difficult to obtain aligned clean and corrupted image pairs. Existing algorithms are all trained with synthetic data. The computationally least expensive synthesis technique is based on the random pixel displacement $+$ blur model \cite{Leonard_Howe_Oxford,Lau2021_sim}. In the optics community, there are techniques based on ray-tracing and wave-propagation \cite{roggemann1996imaging,SchmidtTurbBook,Hardie2017}. A more recent physics-based simulation technique based on the collapsed phase-over-aperture model and the phase-to-space transform is proposed in \cite{chimitt_chan_sim,Mao_2021_ICCV}. Our data synthesis scheme is based on the \textit{P2S} model provided by authors of \cite{Mao_2021_ICCV}.

\section{Restoration Model}

\subsection{Problem Setting and Motivation}
Consider a clean image $\mathbf{I}$ in the object plane that travels through the turbulence to the image plane. Following the classical split-step wave-propagation equation, the resulting image $\widetilde{\mathbf{I}}$ is constructed through a sequence of operations in the phase domain:
\begin{equation}
\mathbf{I} \rightarrow \text{Fresnel} \rightarrow \text{Kolmogorov} \rightarrow \cdots \text{Fresnel} \rightarrow \text{Kolmogorov} \rightarrow \widetilde{\mathbf{I}},
\label{eq: Kolmogorov}
\end{equation}
where ``Fresnel'' represents the wave propagation step by the Fresnel diffraction, and ``Kolmogorov'' represents the phase distortion due to the Kolmogorov power spectral density \cite{Kolmogorov1941}. 

Certainly, Eqn.~\ref{eq: Kolmogorov} is implementable as a forward equation (ie for simulation) but it is nearly impossible to be used for solving an inverse problem. To mitigate this modeling difficulty, one computationally efficient approach is to approximate the turbulence as a composition of two processes:
\begin{equation}
    \widetilde{\mathbf{I}} = \Big( \;\; \underset{\;\; \text{blur}}{\underbrace{\mathcal{H}}}  \;\; \circ \underset{\text{geometric}}{\underbrace{\mathcal{G}}} \Big) (\mathbf{I}) + \mathbf{N},
    \label{eq: main}
\end{equation}
where $\mathcal{H}$ is a convolution matrix representing the spatially \emph{varying} blur, and $\mathcal{G}$ is a mapping representing the geometric pixel displacement (known as the tilt). The variable $\mathbf{N}$ denotes the additive noise / model residue in approximating Eqn.~\ref{eq: Kolmogorov} with a simplified model. The operation ``$\circ$'' means the functional composition. That is, we first apply $\mathcal{G}$ to $\mathbf{I}$ and then apply $\mathcal{H}$ to the resulting image. 

We emphasize that Eqn.~\ref{eq: main} is only a mathematically convenient way to derive an approximated solution for the inverse problem but not the true model. The slackness falls into the fact that the pixel displacement in $\mathcal{G}$ across the field of view are correlated, so do the blurs in $\mathcal{H}$. The specific correlation can be referred to the model construction in the phase space, for example \cite{chimitt_chan_sim}. In the literature, Eqn.~\ref{eq: main} the shortcoming of this model is recognized, although some successful algorithms can still be derived \cite{Anantrasirichai2013,Milanfar2013}.

The simultaneous presence of $\mathcal{H}$ and $\mathcal{G}$ in Eqn.~\ref{eq: main} makes the problem hard. If there is only $\mathcal{H}$, the problem is a simple deblurring. If there is only $\mathcal{G}$, the problem is a simple geometric unwrapping. Generic deep-learning models such as \cite{Yasarla2021ICIP,Nair2021ICIP} adopt network architectures for classical restoration problems based on conventional CNNs, which are developed for one type of distortion. Effective, their models treat the problem as 
\begin{equation}
    \widetilde{\mathbf{I}} = \mathcal{T}(\mathbf{I}) + \mathbf{N},
\end{equation}
where $\mathcal{T} = \mathcal{G}\circ\mathcal{H}$ is the overall turbulence operator. Without looking into how $\mathcal{T}$ is constructed, existing methods directly train a generic restoration network by feeding it with noisy-clean training pairs. Since there is no physics involved in this generic procedure, the generalization is often poor.

Contrary to previous methods, in this paper, we propose to jointly estimate the physical degradation model $\mathcal{T}$ of turbulence along with reconstruction of clean image from the degraded input $\widetilde{\mathbf{I}}$. Such formulation explicitly forces our model to focus on learning a generic turbulence degradation operator independent of image contents, along with the reconstruction operation to generate clean output. Moreover, our network training is assisted by high-quality, large-scale, and physics-motivated synthetic training data to better learn the key characteristics of the atmospheric turbulence effect. The detailed model architecture will be presented in the following subsection.

\begin{figure}
\includegraphics[width=\textwidth, trim=2em 0em 0em 0em]{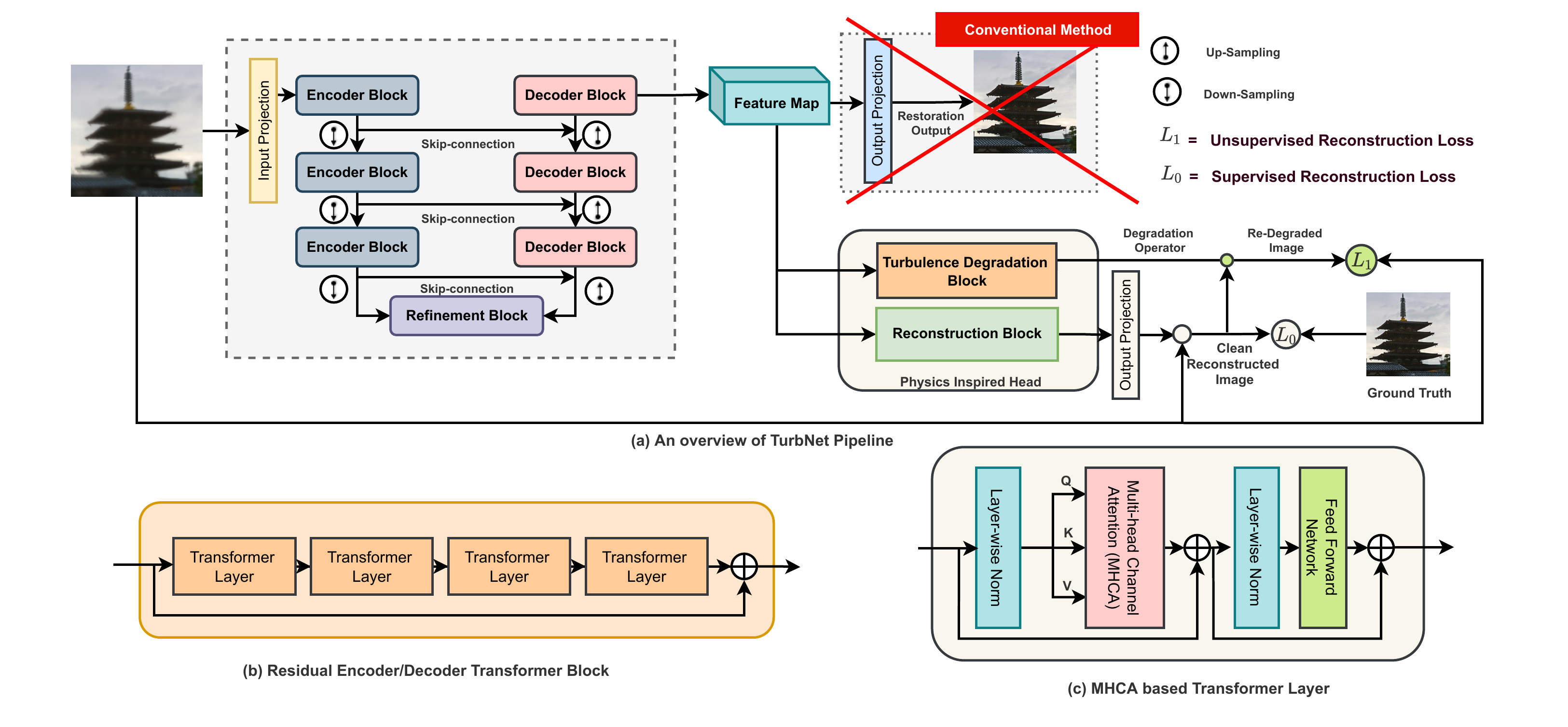}
\caption{\textbf{Architecture of the proposed method}. (a) The overall architecture consists: (i) a transformer to pull the spatially dynamical features from the scene; (ii) instead of directly constructing the image, we introduce a physics-inspired model to estimate the turbulence while reconstructing the image. (b) The structure of the residual encoder/decoder transformer block. (c) The details of each transformer layer.} 
\label{fig:overall_architecture}
\end{figure}

\subsection{Model Architecture}

\paragraph{Turbulence and limitation of CNNs:} CNNs have been \textit{de facto} choice by most of the previous image restoration algorithms, yet they are limited by two primary issues: 1) The convolutional filters cannot adapt to image content during inference due to their static weights. 2) The local receptive fields cannot model the long-range pixel dependencies. A key characteristic of the atmospheric turbulence effect is the ``lucky effect" \cite{Fried78}, meaning that \textbf{image regions} or \textbf{frames} with less degradation will randomly occur due to the distortions being spatially varying. Previous restoration methods treat turbulence restoration as a regression problem using CNNs but ignore the fact that turbulence is highly location adaptive and should not be represented as static fixed kernel applied to all locations. It is not difficult to see that applying static weight convolutions to regions with drastically different distortions will lead to sub-optimal performance. 

The \textit{self-attention} mechanism proposed in recent work \cite{vaswani2017attention,Wang2018NonlocalNN,dosovitskiy2020image} can be a powerful alternative, as it can capture context-dependent global interactions by aggregating information across image regions. Leveraging the capability of multi-head self-attention, we propose the \textbf{\textit{TurbNet}}, a transformer-based end-to-end network for restoring turbulence degraded images. Transformer-based architecture allows the creation of input-adaptive and location-adaptive filtering effect using \textit{key, query,} and \textit{weight}, where \textit{key} and \textit{query} decide content-adaptivity while \textit{weight} brings location-adaptivity. Our design, as shown in Figure \ref{fig:overall_architecture}, is composed of several key building blocks:

\subsubsection{Transformer Backbone:} Our proposed network consists of a transformer-based backbone that has the flexibility of constructing an input-adaptive and location-adaptive unique kernel to model spatially- and instance-varying turbulence effect. Inspired by the success of \cite{ronneberger2015u,wang2021uformer,zamir2021restormer} in various common image restoration tasks (e.g., denoising, deblurring, etc.), TurbNet adopts a U-shape encoder-decoder architecture due to its hierarchical multi-scale representation while remaining computationally efficient. As shown in Figure \ref{fig:overall_architecture} (b),  the residual connection across the encoder-decoder provides an identity-based connection facilitating aggregation of different layers of features. Our backbone consists of three modules: input projection, deep encoder and decoder module. Input project module uses convolution layers to extract low frequency information and induces dose of convolutional inductive bias in early stage and improves representation learning ability of transformer blocks \cite{xiao2021early}. Deep encoder and decoder modules are mainly composed of a sequential cascade of Multi-head channel attention (MHCA) based transformer layers. Compared to prevalent CNN-based turbulence mitigation models, this design allows content-based interactions between image content and attention weights, which can be interpreted as spatially varying convolution \cite{cordonnier2019relationship}.     

The primary challenge of applying  conventional transformer blocks for image restoration task comes from the quadratic growth of key-query dot product interactions, i.e., $\mathcal{O}(W^2H^2)$, for images with $W \times H$ pixels. To alleviate this issue, we adopt the idea of applying self-attention across channels instead of spatial dimension \cite{zamir2021restormer}, and compute cross-covarience across channels generating attention map. Given \textit{query} ($\mathbf{Q}$), \textit{key} ($\mathbf{K}$), and \textit{value} ($\mathbf{V}$), we reshape $\mathbf{Q}$ and $\mathbf{K}$ such that their dot-product generates a transposed-attention map $\mathbf{A} \in \mathbb{R}^{C \times C}$, instead of conventional $\mathbb{R}^{HW \times HW}$ \cite{dosovitskiy2020image}. Overall, the MHCA can be summarized as:
\begin{equation}
    \mathbf{X'} = \mathbf{W_p}\ \text{Attention} (\mathbf{Q, K, V}) + \mathbf{X}
\end{equation}
\begin{equation}
    \text{Attention} (\mathbf{Q, K, V}) = \mathbf{V} \cdot \text{softmax} \bigg\{ \frac{\mathbf{K \cdot Q}}{\alpha}\bigg\}
\end{equation}
where $\mathbf{X'}$ and $\mathbf{X}$ are input and output feature maps, $\mathbf{W_p^{(\cdot)}}$ is the 1 $\times$ 1 point-wise convolution, and $\alpha$ is a learnable scaling parameter to control the magnitude of $(\mathbf{K \cdot Q})$ before applying softmax. 

\subsubsection{Image Reconstruction Block: } To further enhance deep features generated by the transformer backbone, TurbNet uses the reconstruction block. The primary job of the reconstruction block is to take deep features corresponding to turbulence degraded input image $\widetilde{\mathbf{I}}$ by the transformer backbone, further enrich it at high spatial resolution by encoding information from spatially neighboring pixel positions. Next, the enriched features pass through an output projection module with 3 $\times$ 3 convolution layers to project it back low dimension feature map corresponding to the reconstructed clean image $\mathbf{J}$. The design of the reconstruction block is very similar to the encoder block having MHCA, with an introduction of Locally-Enhanced Feed Forward Network (LoFFN) \cite{wang2021uformer}.   

\begin{figure}[h!]
\includegraphics[width=\textwidth, trim = 0em 5em 0em 2em]{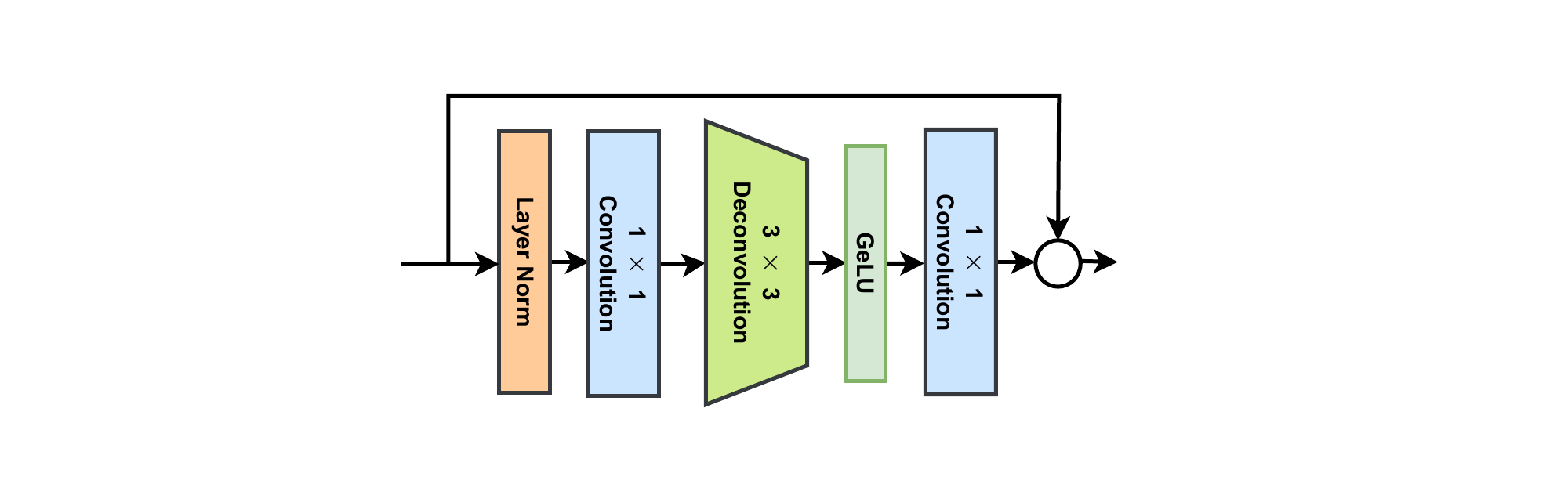}
\caption{Locally-Enhanced Feed Forward Network (LoFFN) used in the image reconstruction block and the turbulence degradation block.} \label{fig:LoFFN}
\end{figure}

Precisely, the work of Reconstruction module can be summarized as:
\begin{equation}
    \underset{\substack{\text{Deep Features of degraded} \\ \text{Input Image $\widetilde{\mathbf{I}}$}}}{\underbrace{\mathbf{F_{\widetilde{I}}}}}
    \rightarrow \text{Reconstruction Module} \rightarrow 
    \underset{\substack{\text{Reconstructed Clean} \\ \text{Output Image}}}{\underbrace{\mathbf{J_{\widetilde{I}}}}}
\end{equation}

\subsubsection{Turbulence Degradation Block:} In TurbNet, the turbulence degradation module learns the physical turbulence degradation operator $\mathcal{T}$ from the input synthetic training data. The primary job of turbulence degradation module is to take clean reconstructed image $\mathbf{J_{\widetilde{I}}}$ corresponding to degraded input image $\mathbf{\widetilde{I}}$, apply the learned degradation operator $\mathcal{T}$, to construct back the \textbf{re-degraded} input image $\mathbf{\widetilde{I}_\mathcal{T}}$. This formulation enriches the training set by incorporating additional latent degradation images ($\mathbf{\widetilde{I}_\mathcal{T}}$), in addition to synthesized degraded images ($\mathbf{\widetilde{I}}$), during the training process. Additionally, this module facilitates self-supervised learning without the availability of ground truth. The architecture of this module is the same as Image Reconstruction Block with LoFFN.

Precisely, the work of Degradation Block can be summarized as:
\begin{equation}
    \underset{\substack{\text{Reconstructed Clean} \\ \text{Output Image}}}{\underbrace{\mathbf{J_{\widetilde{I}}}}}
    \rightarrow \text{Degradation Operator $\mathcal{T(\cdot)}$ } \rightarrow 
    \underset{\substack{\text{Re-degraded} \\ \text{Output Image}}}{\underbrace{\mathbf{\widetilde{I}_\mathcal{T}}}}
\end{equation}

\subsubsection{Loss Function:} TurbNet optimization requires the joint optimization of reconstruction operation and the turbulance degradation operation. Given the synthetic training pair of degraded input $\mathbf{\widetilde{I}}$, and corresponding ground truth image $\mathbf{I}$, we formulate following two losses:
\begin{equation}
     \underset{\text{Supervised Reconstruction Loss}}{\underbrace{\mathcal{L}_0}}\ \ \ \  = ||  \mathbf{J_{\widetilde{I}}} - \mathbf{I} ||_1 
\end{equation}

\begin{equation}
    \underset{\text{Self-supervised Reconstruction Loss}}{\underbrace{\mathcal{L}_1}} = || \mathbf{\widetilde{I}_\mathcal{T}} - \mathbf{\widetilde{I}} ||_1 
\end{equation}
where, $\mathcal{L}_0$ is responsible for constructing a clean image $\mathbf{J_{\widetilde{I}}}$ given the degraded input image $\widetilde{\mathbf{I}}$, $\mathcal{L}_1$ helps to ensure degradation operator $\mathcal{T}$ can reconstruct the original the original input $\widetilde{\mathbf{I}}$ from the reconstructed clean image $\mathbf{J_{\widetilde{I}}}$. 

Eventually, the overall loss $\mathcal{L}$ to train TurbNet can be summarized as:
\begin{equation}
    \mathcal{L} = \alpha \times \mathcal{L}_0 + (1 - \alpha) \times \mathcal{L}_1
\end{equation}

\subsubsection{Overall Pipeline} As shown in Figure \ref{fig:overall_architecture}(a), TurbNet utilizes a U-shape architecture built upon transformer blocks to extract deep image features. As suggested in \cite{xiao2021early}, an initial convolution-based input projection is used to project the input image to higher dimensional feature space, which can lead to more stable optimization and better results. After obtaining the feature maps, TurbNet jointly learns the turbulence degradation operator ($\mathcal{T}$) along with the reconstructed image ($\mathbf{J_{\widetilde{I}}}$), in contrary to general image restoration methods \cite{wang2021uformer,Liu2021SwinTH,Chen2021PreTrainedIP,zamir2021restormer} that directly reconstruct the clean image. This design facilitates spatial adaptivity and long-range dynamics of turbulence effects, plus a self-supervised consistency loss.

\subsubsection{Synthetic-to-Real Generalization:}
With a pre-trained TurbNet model $\mathcal{M(\cdot)}$ using the synthetic data, TurbNet design allows an effective way of generalizing $\mathcal{M(\cdot)}$ on unseen real data (if required) with the help of degradation operator $\mathcal{T}(\cdot)$ in a self-supervised way. Starting from model $\mathcal{M(\cdot)}$, we create a generalization dataset by incorporating unlabelled real data with the synthetic data to fine-tune $\mathcal{M(\cdot)}$. For input images with no ground truth, $\mathcal{M(\cdot)}$ is optimized using Equation (9), while for input images from labeled synthetic data $\mathcal{M(\cdot)}$ is optimized using Equation (8, and 9). Note that we incorporate synthetic data into the fine-tuning process to mitigate the issue of catastrophic forgetting during generalization.

\section{Large-Scale Training and Testing Datasets}

\subsection{Training Data: Synthetic Data Generating Scheme}

Training a deep neural network requires data, but the real clean-noisy pair of turbulence is nearly impossible to collect. A more feasible approach here is to leverage a powerful turbulence simulator to synthesize the turbulence effects. 

\begin{figure}
\centering
\includegraphics[width=7cm]{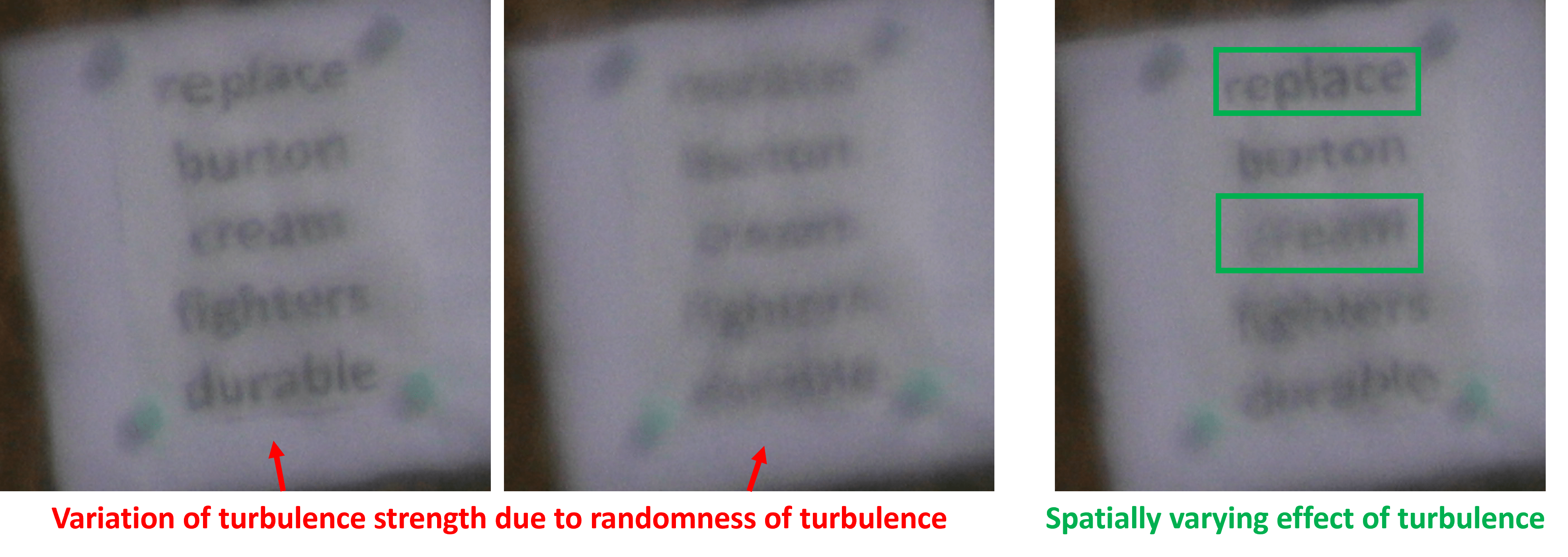}
\caption{Key turbulence effects requiring attention while designing synthetic dataset. } 
\label{fig:illustration}
\end{figure}

Turbulence simulation in the context of deep learning has been reported in \cite{Yasarla2021ICIP,Nair2021ICIP,Lau2021}. Their model generates the geometric distortions by repeatedly smoothing a set of random spikes, and the blur is assumed to be spatially invariant Gaussian \cite{Lau2021_sim}. We argue that for the face images studied in \cite{Yasarla2021ICIP,Nair2021ICIP,Lau2021}, the narrow field of view makes their simplified model valid. However, for more complex scenarios, such a simplified model will fail to capture two key phenomena that could cause the training of the network to fail: (1) The instantaneous distortion of the turbulence can vary significantly from one observation to another even if the turbulence parameters are fixed. See Figure~\ref{fig:illustration}(a) for an illustration from a real data. (2) Within the same image, the distortions are spatially varying. See Figure~\ref{fig:illustration}(b). 

In order to capture these phenomena, we adopt an advanced simulator \cite{Mao_2021_ICCV} to synthesize a large-scale \emph{training} dataset for atmospheric turbulence. The clean data used by the simulator is the \textit{Places} dataset \cite{zhou2017places}. A total of 50,000 images are generated, and the turbulence parameters are configured to cover a wide range of conditions. The details of the simulation can be found in the supplementary material. We remark that this is the first attempt in the literature to systematically generate such a comprehensive and large-scale training dataset.

\subsection{Testing Data: Heat Chamber and Text Datasets}

Our real benchmarking dataset consists of two parts: the \textit{Heat Chamber Dataset} and the \textit{Turbulent Text Dataset}. Although this paper focuses on single frame restoration, both our benchmarking datasets contain 100 static turbulence degraded frames for each scene. We believe that by doing so, researchers in the field working on multi-frame reconstruction can also benefit from our dataset. Both datasets will be made \textbf{publicly available}.

\begin{figure}
\centering
\includegraphics[width=0.8\textwidth, trim=0em 4em 0em 4em]{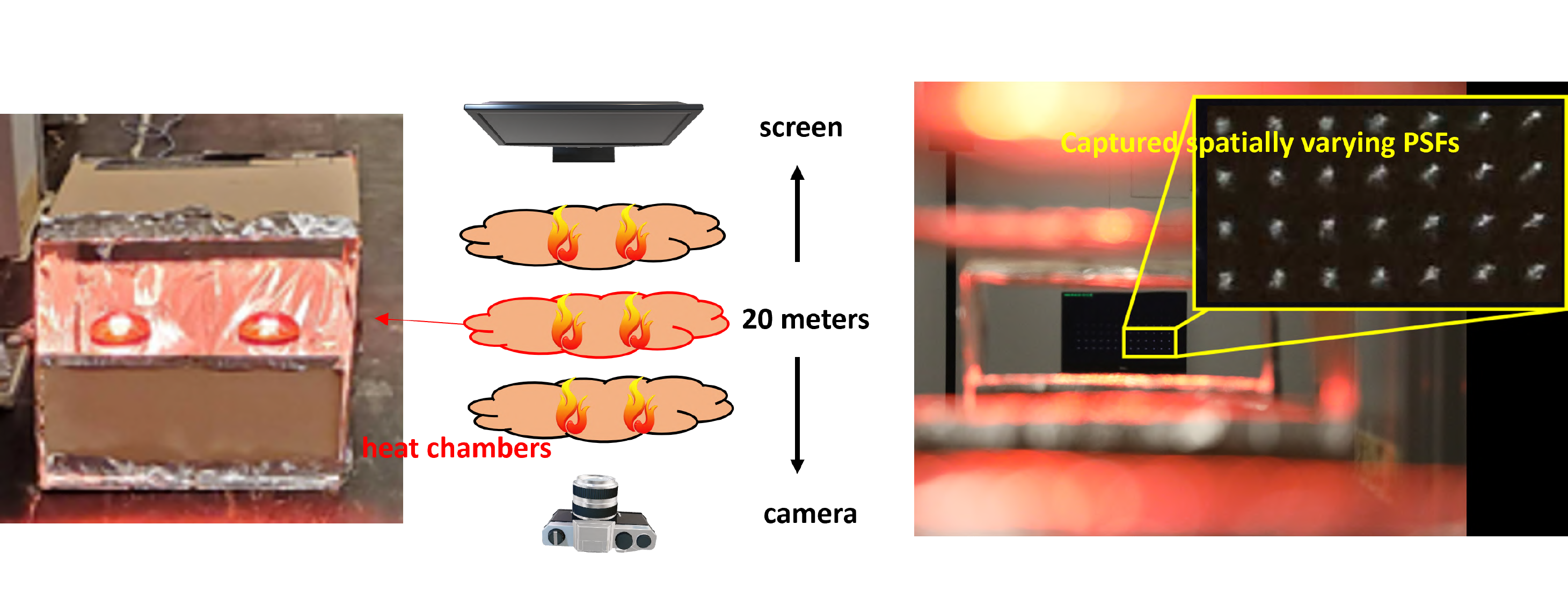}
\caption{The setup of heat chamber data collection. We evenly placed three heat chambers along the imaging path. Our dataset captures better spatially varying effect. }
\label{fig:heat_setup}
\end{figure}

\textbf{Heat Chamber Dataset. } The \textit{Heat Chamber Dataset} is collected by heating the air along the imaging path to artificially create a stronger turbulence effect. The setup for collecting the heat chamber dataset is shown in \ref{fig:heat_setup}. Turbulence-free ground truth images can be obtained by shutting down the heat source. The images are displayed on a screen placed 20 meters away from the camera. 

\begin{figure}[h!]
	\centering
	\begin{tabular}{c c c c c c}
		\includegraphics[width=0.12\textwidth]{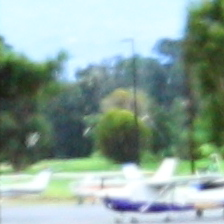}&
		\includegraphics[width=0.12\textwidth]{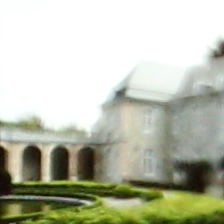}&
		\includegraphics[width=0.12\textwidth]{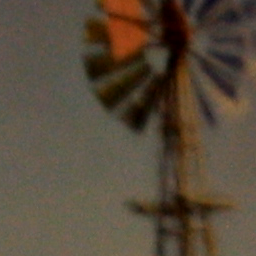}&
		\includegraphics[width=0.12\textwidth]{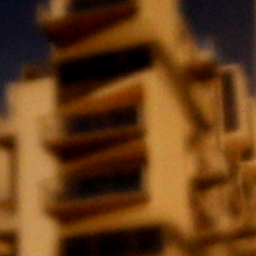}&
		\includegraphics[width=0.12\textwidth]{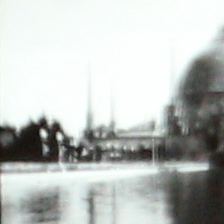}&
		\includegraphics[width=0.12\textwidth]{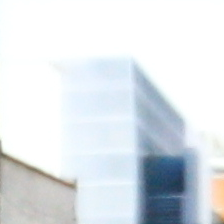}\\
		
		\includegraphics[width=0.12\textwidth]{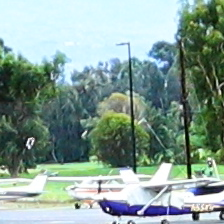}&
		\includegraphics[width=0.12\textwidth]{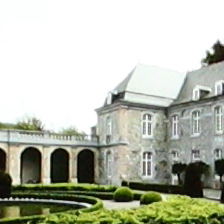}&
		\includegraphics[width=0.12\textwidth]{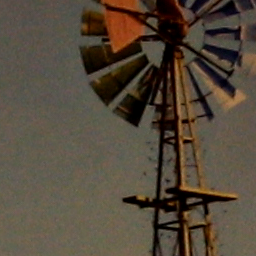}&
		\includegraphics[width=0.12\textwidth]{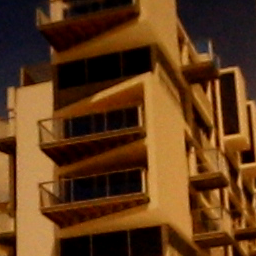}&
		\includegraphics[width=0.12\textwidth]{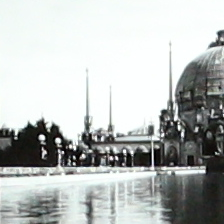}&
		\includegraphics[width=0.12\textwidth]{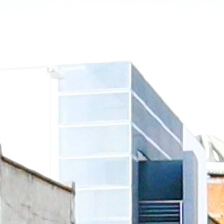}\\
	\end{tabular}
	\caption{Sample turbulence degraded images (top) and corresponding ground truth (bottom) from our \textit{Heat Chamber Dataset}. The $D/r_0$ is estimated to be around 3. }
	\label{fig:heat_sample}
\end{figure}

We remark that while similar datasets have been collected in  \cite{Hirsch2010,Anantrasirichai2013}, our data has a clear improvement: we use a long path and more evenly distributed heat so that the turbulence effect is closer to the true long-range effect. The captured images have a better anisoplanatic (spatially varying) effect such that an almost distortion-free frame is less likely to occur compared with the dataset in \cite{Hirsch2010,Anantrasirichai2013}. In addition, our dataset is much large in scale. It contains 2400 different images, which allows for a better evaluation of the learning-based model. Sample images of the \textit{Heat Chamber Dataset} can be found in Figure \ref{fig:heat_sample}. 

\textbf{Turbulence Text Dataset. } Due to the nature of the problem, it is extremely difficult, if not impossible, to capture ground truth clean images in truly long-range settings. Therefore, we adopt the idea of using the performance of high-level vision task as an evaluation metric for image restoration \cite{Li2019dehaze,ICCV17aodnet}. Specifically, we calculate the detection ratio and longest common subsequence on the output of an OCR algorithm \cite{tian2016detecting,OCR1} as the evaluation metrics. The terms will be defined in section 5.4. 

There are several advantages of using text recognition: 1) The degradation induced by atmospheric turbulence, the geometric distortion and the loss of resolution, can be directly reflected by the text patterns. Both types of degradation need to be removed for the OCR algorithms to perform well. 2) The OCR is a mature application. The selected algorithms should be able to recognize the text patterns as long as the turbulence is removed. Other factors such as the domain gap between the training and testing data will not affect the evaluation procedure as much as other high-level vision tasks. 3) An important factor to consider when designing the dataset is whether the difficulty of the task is appropriate. The dataset should neither be too difficult such that the recognition rate cannot be improved by the restoration algorithms nor too easy making all algorithms perform similarly. We can easily adjust the font size and contrast of text patterns to obtain a proper difficulty level. 

\begin{figure}[h!]
\centering
\includegraphics[width=10cm]{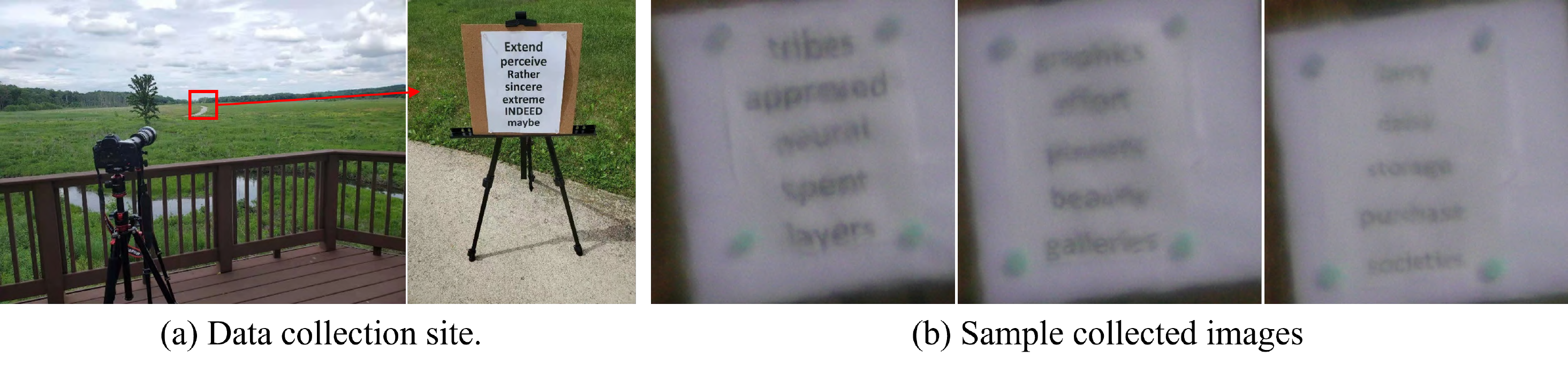}
\caption{Data collection site of the \textit{Turbulence Text Dataset}. The distance between the camera and the target is 300 meters. The $D/r_0$ is estimated to be in range of 2.5 to 4 (varies due to the temperature change during the collection process). The collected text patterns are in 3 different scales. } 
\label{fig:text_samples}
\end{figure}

The \textit{Turbulence Text Dataset} consists of 100 scenes, where each scene contains 5 text sequences. Each scene has 100 static frames. It can be assumed that there is no camera and object motion within the scene, and the observed blur is caused by atmospheric turbulence. The text patterns come in three different scales, which adds variety to the dataset. We also provide labels to crop the individual text patterns from the images. Sample images from the dataset are shown in Figure \ref{fig:text_samples}.

\begin{table}[h!]
\centering
\small
\caption{Performance comparison of state-of-art restoration baselines with respect to TurbNet on synthetic and \textit{Heat Chamber} dataset.}
\begin{tabular}{lcccccc}
\toprule
 & TDRN\cite{yasarla2020learning} & MTRNN\cite{park2020multi} & MPRNet\cite{zamir2021multi} & Uformer\cite{wang2021uformer} & Restormer\cite{zamir2021restormer}  & \textbf{TurbNet}\\
\midrule
&\multicolumn{6}{c}{\textbf{Synthetic Dataset}}\\
\midrule
PSNR & 21.35 & 21.95 & 21.78 & 22.03 & 22.29 & \textbf{22.76}\\
SSIM & 0.6228 & 0.6384 & 0.6410 & 0.6686 & 0.6719 & \textbf{0.6842}\\
\midrule
&\multicolumn{6}{c}{\textbf{HeatChamber Dataset}}\\
\midrule
PSNR & 18.42 & 18.12 & 18.68 & 19.12 & 19.01 & \textbf{19.76}\\
SSIM & 0.6424 & 0.6379 & 0.6577 & 0.6840 & 0.6857 & \textbf{0.6934}\\
\bottomrule
\end{tabular}
\label{table:syn_psnr_ssim}
\end{table} 

\section{Experiment Results}

\subsubsection{Implementation Details:} TurbNet uses a 4-staged symmetric encoder-decoder architecture, where stage 1, 2, 3, and 4 consist of 4, 6, 6, and 8 MHCA-based transformer layers respectively. Our Reconstruction block and Turbulence Degradation block consist of 4 MHCA-transformer layers enhanced with LoFFN. TurbNet is trained using $50,000$ synthetic dataset generated using a physics-based stimulator \cite{Mao_2021_ICCV} and MIT Places dataset \cite{zhou2017places} while synthetic evaluation results are generated on $5,000$ synthetic images. Due to resource constraint, our synthetic training uses a batch size of 8 with Adam optimizer. We start our training with learning rate of $1e-4$, and use the cosine annealing scheduler to gradually decrease the learning rate over the span of 50 epcohs. During training, to modulate between the loss Equation 8 and 9, we have use $\alpha$ to be 0.9. All the baselines method used in our evaluation has been trained with exactly same settings and same dataset using their official GitHub implementation for fair comparison. Additional implementation details are provided in supplementary materials.

\subsection{Synthetic and \textit{Heat Chamber} Dataset Results}

\begin{figure}[h!]
\centering
\includegraphics[width=11cm, trim=2em 136em 0em 155em]{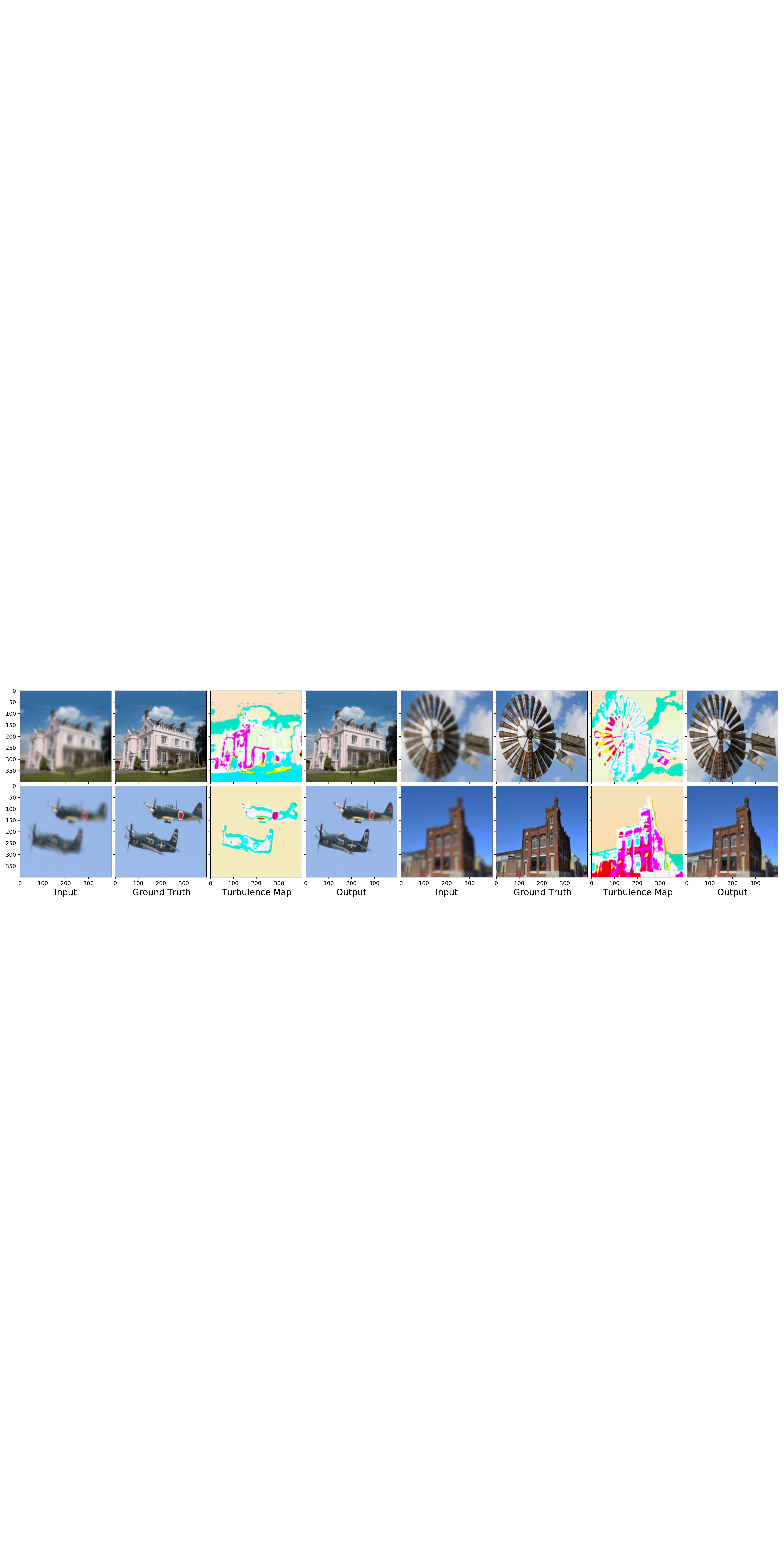}
\caption{Qualitative Performance comparison of TurbNet wrt. the ground truth. } \label{fig:synthetic_examples}
\end{figure}

We first conduct an experiment on a synthetic testing dataset generated with the same distribution as testing data. In Figure \ref{fig:synthetic_examples}, we show a qualitative comparison between our restored images with ground truth. It can be seen that our results are accurately reconstructed with the assist from estimated turbulence map. 

\begin{figure}[h!]
\centering
\includegraphics[width=11.5cm, trim=0em 132em 0em 135em]{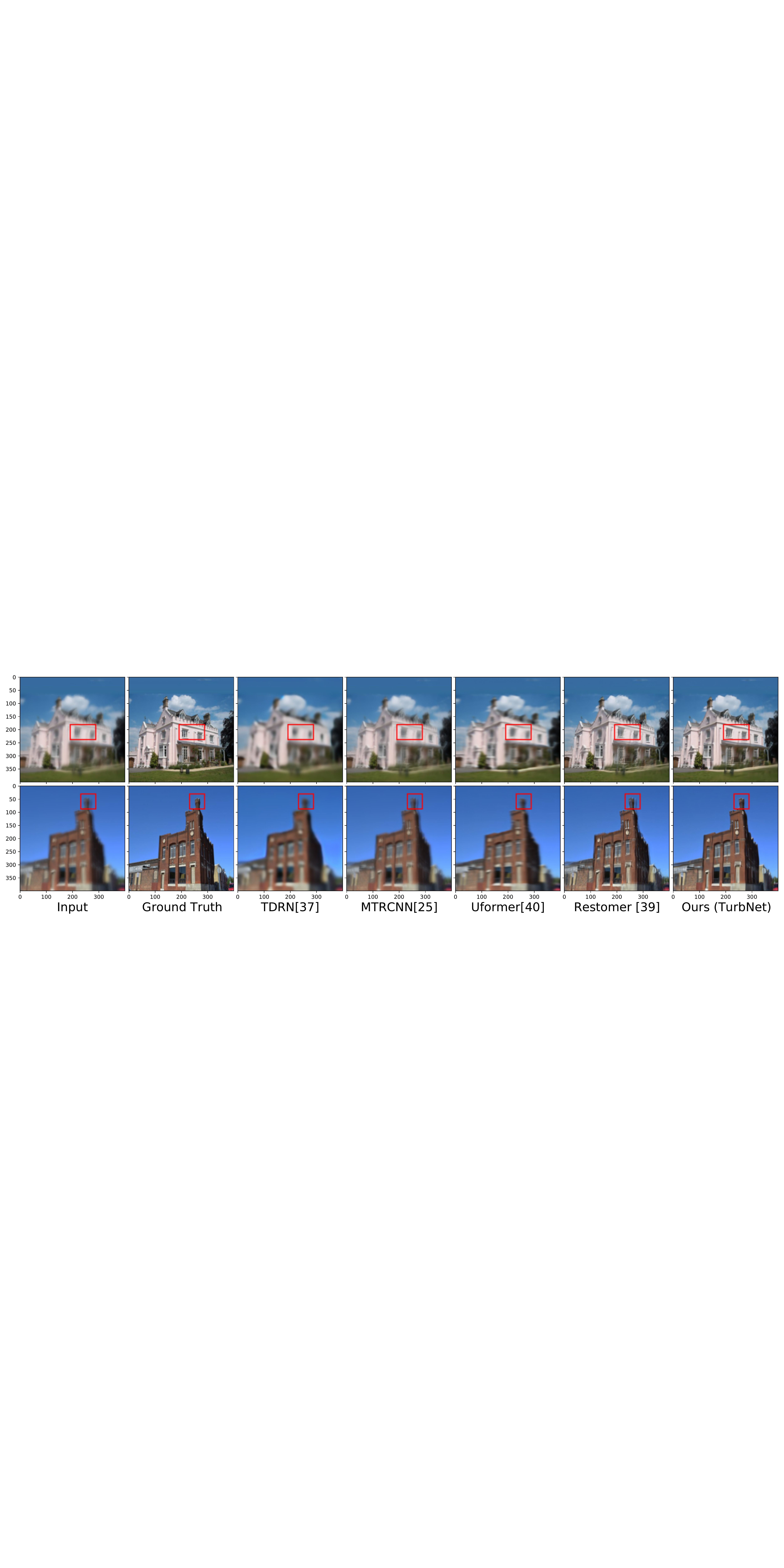}
\caption{Qualitative Performance comparison of TurbNet wrt. other SOTA methods. }
\label{fig:method_comparison}
\end{figure}

We then compare our results qualitatively with the existing algorithms on both synthetic and \textit{Heat Chamber} dataset. A Visual comparison on the synthetic dataset can be found in Figure \ref{fig:method_comparison}. It can be observed that the transformer-based methods generally perform better than the CNN-based methods due to their ability to adapt dynamically to the distortions. The proposed method achieves authentic reconstruction due to its ability to explicitly model the atmospheric turbulence distortion. Table \ref{table:syn_psnr_ssim} presents the quantitative evaluation of TurbNet wrt. other baselines. TurbNet achieves the best results in both PSNR and SSIM. Note that Uformer\cite{wang2021uformer}, and Restomer\cite{zamir2021restormer} (designed for classical restoration problems like deblurring, deraining, etc.) uses  transformer-based encoder decoder architecture, but their performance is significantly low than TurbNet, which validates the importance of our decoupled (reconstruction and degradation estimation) design.

\subsection{\textit{Turbulence Text Dataset} Results}

\subsubsection{Evaluation Method:} In order to evaluate the performance of TurbNet on our real-world turbulence text dataset, we use publicly available OCR detection and recognition algorithms \cite{tian2016detecting,OCR1}. We propose the following two evaluation metrics - Average Word Detection Ratio ($\mathbf{AWDR}$), and Average Detected Longest Common Subsequence ($\mathbf{AD-LCS}$) defined as follows:
\begin{equation}
    \mathbf{AWDR} = \frac{\sum_{Scene = 1}^N\frac{\text{Word Detected}_{scene}}{\text{Word Count}_{scene}}}{N}, 
\end{equation}

\begin{equation}
   \mathbf{AD-LCS} = \frac{\sum_{Scene = 1}^N \sum_{Word = 1}^K \mathcal{LCS}(Detected String, True String)}{N}, 
\end{equation}
where $\mathcal{LCS}$ represents the Longest Common Subsequence, $True String$ represents the ground truth sequence of characters corresponding to a word $i$ in the image, $Detected String$ represents a sequence of characters recognized by OCR algorithms for word $i$, and $N$ is the total number of scenes in the test dataset.

\begin{table}[h!]
\centering
\small
\caption{Performance comparison of state-of-art restoration baselines with respect to TurbNet on our \textit{Turbulence Text Dataset}.}
\begin{tabular}{lcccccc}
\toprule
 & Raw Input & TDRN\cite{yasarla2020learning} & MTRNN\cite{park2020multi} & MPRNet\cite{zamir2021multi} &  Restormer\cite{zamir2021restormer}  & \textbf{TurbNet}\\
\midrule
AWDR & 0.623 & 0.617 & 0.642 & 0.633 & 0.702 & \textbf{0.758}\\
AD-LCS & 5.076 & 5.011 & 5.609 & 5.374 & 6.226 & \textbf{7.314}\\
\bottomrule
\end{tabular}
\label{table:ocr_challenege}
\end{table} 

\subsubsection{Discussion:} 
\begin{figure}[h!]
\centering
\includegraphics[width=7.5cm, trim=2em 40em 0em 40em]{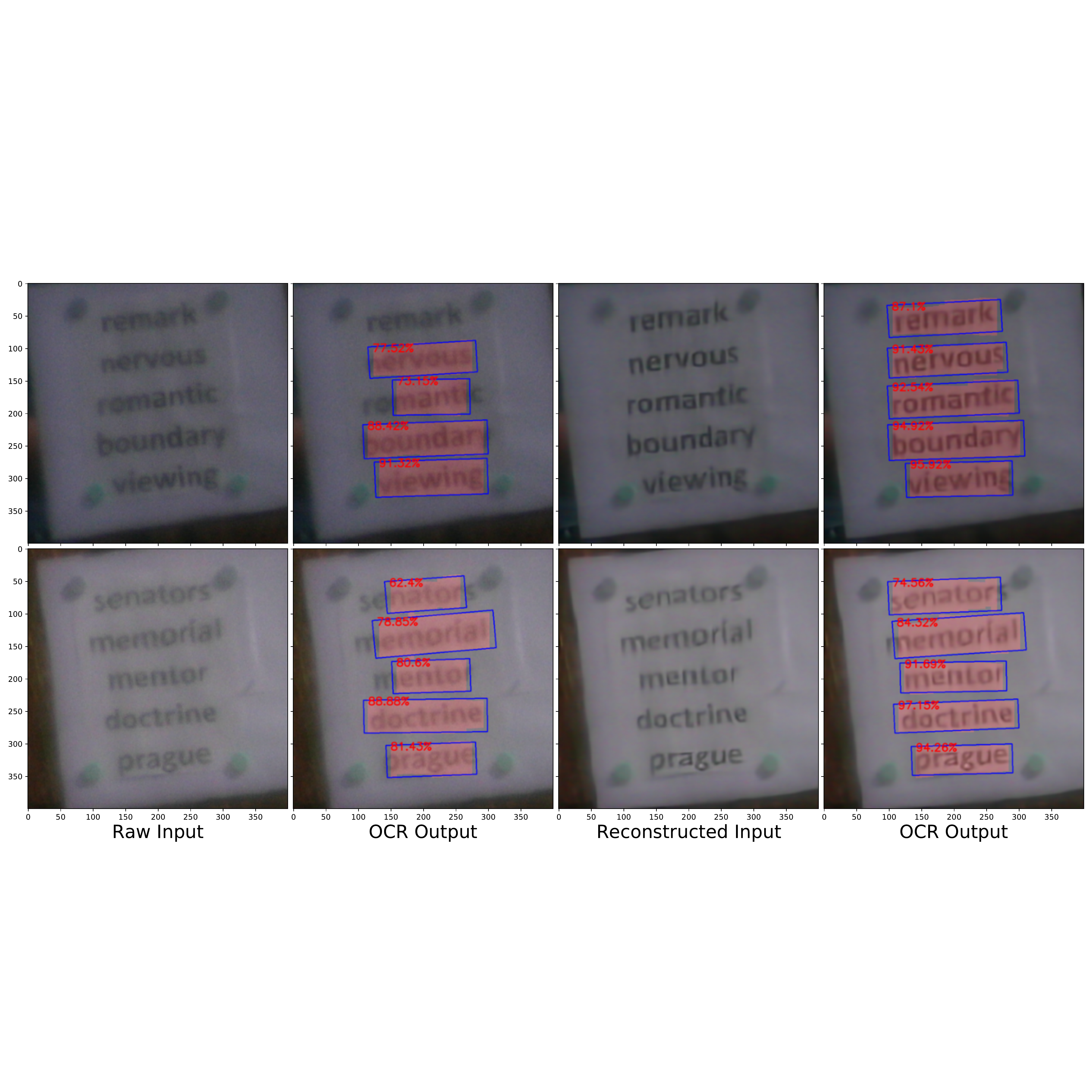}
\caption{OCR performance of our reconstruction algorithm for  \textit{Turbulance Text Dataset} } \label{fig:ocr_comparison}
\end{figure}

Figure \ref{fig:ocr_comparison} represents the performance of OCR on the real turbulence impacted images and images restored by TurbNet. It is evident that our restoration model significantly helps in improving the OCR performance by identifying comparatively more words with higher confidence.  Table \ref{table:ocr_challenege} presents the performance gain by TurbNet over the real turbulence degraded text images and their restored version by various state-of-the-art methods. OCR algorithms achieve massive improvements of +0.135 (AWDR) and +2.238 (AD-LCS) when used on images restored by TurbNet compared to being used directly on real images from our proposed test dataset.

\subsection{Experimental Validity of the Proposed Model} 
\begin{figure}[h!]
	\centering
	\includegraphics[width=0.8\textwidth,trim = 0em 1em 0em 1em]{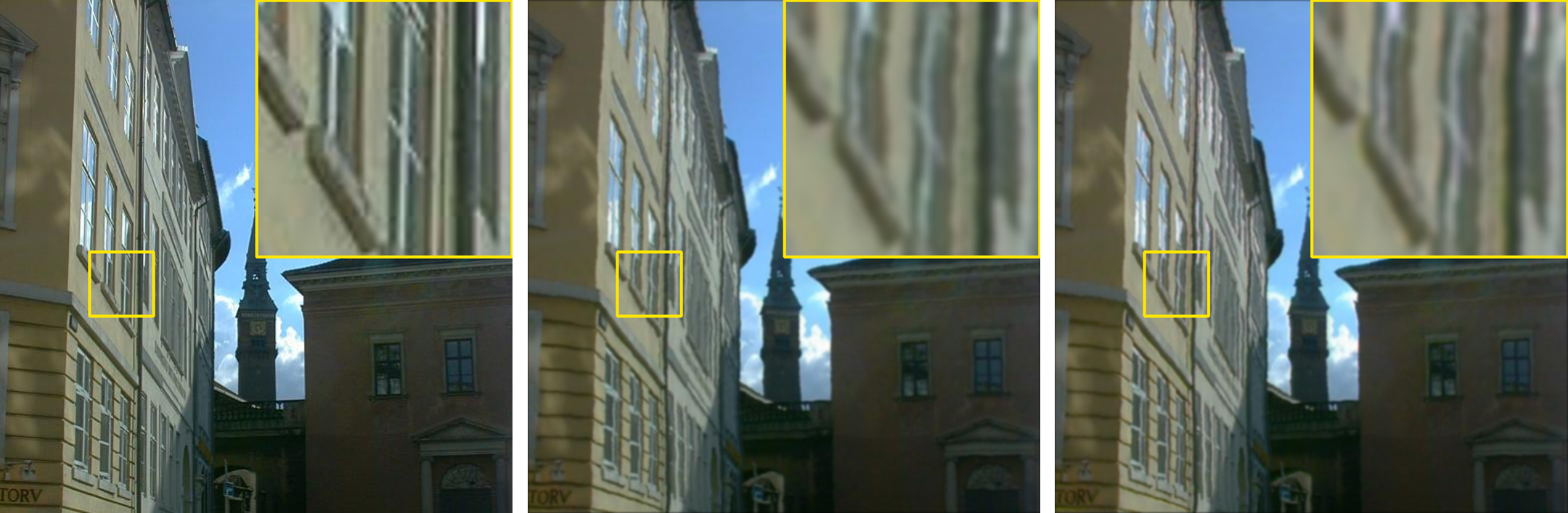}
	\caption{Validation of our turbulence map. Left: groundtruth. Middle: original turbulence frame. Right: groundtruth re-corrupted with the extracted turbulence map.}
	\label{fig:vis}
\end{figure}

We conduct two additional experiments to validate the proposed model. The first experiment is an ablation study, where we demonstrate the impact of replacing transformer as feature encoder with U-Net \cite{ronneberger2015u} and removing the turbulence map estimation part. The result is reported in \ref{tab:ablation}, where we observe a significant performance drop in both cases. The second experiment is to prove the effectiveness of the extracted turbulence map. We extract a turbulence map from a simulated frame and apply the map back to the ground-truth image. We calculate the PSNR of this re-corrupted image w.r.t. the original turbulence frame. We tested on 10K turbulence frames and the average PSNR is \textbf{39.89 dB}, which is a strong evidence that our turbulence map can effectively extract the turbulence information embedded in the distorted frames. A visualization of the experiment can be found in Figure \ref{fig:vis}. 

\begin{table}[h]
    \small
    \centering
    \caption{Ablation on Heat Chamber Dataset}
    \begin{tabular}{c|cc}
    \toprule
        Model type & PSNR & SSIM \\
    \midrule
        \textbf{TurbNet [Ours]} & 19.76 & 0.6934\\
        TurbNet - Turbulance Map & 19.03 ($\downarrow$) & 0.6852 ($\downarrow$)\\
        TurbNet - Transformer & 18.62  ($\downarrow$) & 0.6481 ($\downarrow$)\\
    \bottomrule
    \end{tabular}

    \label{tab:ablation}
\end{table} 

\section{Conclusions} 
In this work, identifying the short-come of existing image restoration algorithms, we propose a novel physics-inspired turbulence restoration model (TurbNet) based on transformer architecture to model spatial adaptivity and long-term dynamics of turbulence effect. We present a synthetic data generation scheme for tuning a
sophisticated physics-grounded simulator to generate a large-scale dataset,
covering a broad variety of atmospheric turbulence effects. Additionally, we introduce two new large-scale testing datasets that allow for evaluation with classical objective metrics and a new task-driven metric with optical text recognition. Our comprehensive evaluation on realistic and diverse datasets leads to exposing limitations of existing methods and the effectiveness of TurbNet.

\section*{Acknowledgement}

The research is based upon work supported in part by the Intelligence Advanced Research Projects Activity (IARPA) under Contract No. 2022‐21102100004, and in part by the National Science Foundation under the grants CCSS-2030570 and IIS-2133032. The views and conclusions contained herein are those of the authors and should not be interpreted as necessarily representing the official policies, either expressed or implied, of IARPA, or the U.S. Government. The U.S. Government is authorized to reproduce and distribute reprints for governmental purposes notwithstanding any copyright annotation therein. 

\par\vfill\par

\clearpage
%
%
\bibliographystyle{splncs04}
\bibliography{egbib}
\end{document}